\documentclass[preprint,showpacs,floats,nofootinbib,superscriptaddress,
  showkeys]{revtex4}

\usepackage{bm}
\usepackage{amsmath}
\usepackage{graphicx}

\newcommand{\dd}{{\mathrm{d}}}

\begin{document}

\title{Role of 2p-2h MEC excitations in superscaling }

\author{A. De Pace}
\author{M. Nardi}
\author{W. M. Alberico}
\affiliation{Istituto Nazionale di Fisica Nucleare, Sezione di Torino and
  Dipartimento di Fisica Teorica, via Giuria 1, I-10125 Torino, Italy}
\author{T. W. Donnelly}
\affiliation{Center for Theoretical Physics, Laboratory for 
  Nuclear Science  and Department of Physics, Massachusetts 
  Institute of Technology, Cambridge, MA 02139, USA}
\author{A. Molinari}
\affiliation{Istituto Nazionale di Fisica Nucleare, Sezione di Torino and
  Dipartimento di Fisica Teorica, via Giuria 1, I-10125 Torino, Italy}

\begin{abstract}

Following recent studies of inclusive electron scattering from
nuclei at high energies which focused on two-nucleon emission 
mediated by meson-exchange currents, in this work the superscaling
behavior of such contributions is investigated. Comparisons are
made with existing data below the quasielastic peak where at high 
momentum transfers scaling of the second kind is known to be 
excellent and scaling of the first kind is good, in the proximity
of the peak where both 1p-1h and 2p-2h contributions come into play, 
and above the peak where inelasticity becomes important and one finds
scaling violations of the two kinds.

\end{abstract}

\maketitle

\section{Introduction}

The superscaling behavior of the nuclear inclusive semi-leptonic 
electroweak response in the quasielastic (QE) region has been 
seen to be a useful concept, as illustrated and 
discussed in recent work~\cite{Don99-1,Don99-2,Mai02}. Superscaling 
stands for scaling both in the three-momentum $q\equiv|\bf{q}|$ 
transferred from the probe to the nucleus (scaling of the first 
kind, often referred to as $y$-scaling) and in a momentum that 
is characteristic of a given nucleus (scaling of the second kind). 
The latter characteristic parameter is often taken for simplicity 
to be the Fermi momentum $k_F$, and hence equivalently scaling 
of the second kind is related to the density ($\propto k_F^3$) of 
a given nuclear species. Of course, the choice of $k_F$ as a scaling 
variable arises naturally in the Relativistic Fermi Gas (RFG) model 
which was used initially in Ref.~\cite{Alb88} to motivate the 
introduction of the concept of superscaling. 

In particular, in studies of scaling the inclusive cross sections (or, if
available, the individual separated responses) are first divided by 
an appropriate single-nucleon cross section. For example, in electron 
scattering from a nucleus with charge and neutron numbers $Z$ and $N$, 
respectively, the dividing factor has the form 
$Z\sigma_{ep}+N\sigma_{en}$. For scaling of the first kind one 
displays this reduced cross section against an appropriately chosen 
scaling variable, such as the familiar $y$ variable or a dimensionless 
analog called $\psi$ (or $\psi'$ if a phenomenological energy shift is 
included; see the discussions in the next section for details), and
in some kinematic regimes observes a universality of the results with 
respect to the choice of momentum transfer. That is, the result scales. 
Specifically one finds that this behavior occurs when the momentum 
transfer is sufficiently large (typically larger than about 500-700 
MeV/c) and only when the energy transfer is lower than the value that 
characterizes the position of the quasielastic peak (QEP), the 
so-called scaling region. In contrast, as emphasized in 
Refs.~\cite{Don99-1,Don99-2,Mai02} and also in 
Refs.~\cite{Bar98,Cen97}, scaling of the second kind proceeds from 
the same reduced cross section, but now (1) makes the reduced
cross section or scaling function dimensionless 
by multiplying it by the characteristic nuclear momentum scale 
(called $k_F$ henceforth) and (2) displays this as a function of 
a dimensionless scaling variable, again scaled by division by $k_F$. 
In the scaling region one finds universality for different nuclear 
species at given kinematics. When both types of scaling occur one 
says that the response superscales.

In fact, only the longitudinal part of the response appears to 
superscale (see Ref.~\cite{Jou96} and also 
\cite{Don99-1,Don99-2,Mai02}), and even in the scaling region 
one finds some degree of scaling violation which appears to arise 
in the transverse part of the response. Thus the 
longitudinal contributions, apparently being essentially impulsive 
at high energies, are usually used to determine the basic nuclear 
physics of quasielastic scattering. This leaves us with the transverse 
contributions to explain. In the region above the QEP it is natural 
to have scaling violations, since the reaction mechanism there is 
not solely impulsive knockout of protons and neutrons, but may 
proceed via meson production including via baryon resonances such 
as the $\Delta$. One knows that these contributions are much more 
prominent in the tranverse response that in the longitudinal 
contributions and hence it is reasonable to expect some of 
the scaling violations to arise this way. But this is not the 
entire story: even below meson production threshold there are 
scaling violations in the transverse response. One source for 
this is expected to be the role played by the Meson-Exchange 
Current (MEC) contributions, which again are predominantly 
transverse (see for example Ref.~\cite{Alb90}), and this has 
motivated the present study. 

The correlation and MEC contributions to quasielastic and $\Delta$-region
electron scattering have been the subject of many studies (see, for instance,
\cite{Don78,Van81,Koh81,Alb84,Blu89,Lei90,Cio90,Ama92b,Dek94,Car94,Van95,Ang96,Fab97,Gil97,Gad98,Bau00,Car02,Meu02}
and references therein). 
In this work we shall present a first exploration 
of the impact of the two~particle--two~hole (2p-2h) excitations 
induced by the MEC on the superscaling behavior of the transverse 
RFG response. In the present work we limit the scope to MEC 
contributions mediated by pion exchange and include contributions 
both from intermediate nucleons and also $\Delta$ isobars. 
Our results are compared with the relevant data to assess the 
role of such MEC effects in explaining the scale breaking.

In carrying out this task we take advantage of a very recent 
calculation of the 2p-2h MEC contribution to the RFG transverse 
response performed fully respecting Lorentz and translational 
invariance \cite{DeP03}. The same type of analysis of the RFG 
response has also been carried out in the 1p-1h sector, for both the 
longitudinal and transverse channels, in a scheme that is not only
Lorentz and translational, but gauge invariant as well \cite{Ama02a}.
In order to fulfill gauge invariance also in the 2p-2h sector of 
the RFG Hilbert space, an extension to include 2p-2h 
excitations induced by correlations, in addition to the MEC, is 
required. This topic is currently under investigation. 

In the present paper we start by briefly outlining in Sect.~\ref{sec:form}
the formalism that lies at the basis of our study.
Next, in Sect.~\ref{sec:psi<0}, we explore the kinematical domain 
to the left of the QEP (that is, $\psi'\le0$), which is dominated, 
for $\psi'>-1$, by 1p-1h excitations. Past 
investigations~\cite{Don99-1,Don99-2,Mai02} have convincingly 
demonstrated that for $\psi'\le0$ second-kind scaling is quite well 
satisfied: indeed, the data scale almost perfectly in $k_F$ if the 
latter is appropriately chosen\footnote{It thus appears that this 
offers a third method, beyond the ones based upon the nuclear 
density and the width at half-height of the QEP, to extract this 
crucial parameter of nuclear structure from the data.}. Also, at least
at sufficiently large values of $q$,  
scaling of the first kind appears quite well obeyed by nature for 
$\psi'\le0$, although not as well as scaling of the second kind. 

In this work we observe an important contribution of 2p-2h MEC in the
kinematical region lying below $\psi'=-1$, where the RFG 1p-1h 
processes are forbidden. In the region nearer the QEP where the 
RFG 1p-1h contributions enter for the transverse scaling function 
one has, on the one hand, both the 1p-1h MEC and correlation 
contributions~\cite{Ama02a,Ama02b,Ama03} which conspire and 
interfere with the impulsive contributions (the basic RFG 
responses) to yield a net transverse response that is reduced 
with respect to the RFG. On 
the other hand, the 2p-2h MEC contributions add incoherently and 
tend to compensate the above reduction. See also 
Refs.~\cite{Ama92,Ama93,Ama94}. Note that both classes of 
contributions provide scaling violations. 

After exploring the scaling region and the region of the QEP, in 
Sect.~\ref{sec:psi>0} we extend our study to the domain $\psi'\ge0$, 
focusing on the impact of the 2p-2h excitations on scaling in the
resonance region. Here first-kind scaling is well-known to be very badly
violated, owing to the role now played by nucleon excitations and 
meson production, which, as stated above, are largely transverse.
On the other hand, in Ref.~\cite{Bar04} it is shown that violations of the
transverse second-kind scaling behavior are not so large, although 
they can be as large as 20\% in the resonance region. Again we 
emphasize that in the present work longitudinal scaling is always 
assumed to be perfect. 

In Sect.~\ref{sec:Delta} we explore the role of the $\Delta$ in the 2p-2h
MEC contributions relative to effects arising from two-body currents
that contain only nucleons and pions. Some flexibility exists in
modeling the former via the effective Lagrangian used in the present
work and we present results displaying the impact of changing the
parameters in the model.

Clearly, part of the story is contained in the present work where 
the 2p-2h MEC and their impact on the two kinds of scaling is explored; 
however, naturally the complete story will have to await the 
ultimate extensions to larger classes of MEC and correlation 
effects which are presently being pursued. In the Conclusions we 
address this issue of what is still missing in our analysis of 
superscaling and indicate what is being done to complete the 
(highly non-trivial) study. We also briefly summarize our present 
findings and comment on their significance and on future perspectives.


\section{Formalism}
\label{sec:form}

In this section we summarize the basic formulas needed in our 
analysis of the superscaling physics, closely following 
Refs.~\cite{Don99-1,Don99-2,Mai02}.
As is well-known, the inclusive cross section for the scattering of
electrons from nuclei in the one-photon exchange approximation reads
\begin{equation}
  \frac{\dd^2\sigma}{\dd \Omega_e \dd \omega} =
  \sigma_M \left[ v_L R_L(\kappa,\lambda)+v_TR_T(\kappa,\lambda) \right],
\label{Xsec}
\end{equation}
where $\sigma_M$ is the Mott cross section and $Q^\mu=(\omega,\bm{q})$ 
is the four-momentum transferred by the virtual photon, with $\omega$ 
its energy transfer, $\bm{q}$ its three-momentum transfer and 
$Q^2=\omega^2-q^2$. 
Equivalently we may use the dimensionless variables 
$\lambda\equiv \omega/2m_N$ and $\kappa\equiv |\bm{q}|/2m_N$, 
$ m_N $ being the nucleon mass. One then has for the square of 
the four-momentum transfer 
$|Q^2|/4m_N^2 \equiv \tau = \kappa^2-\lambda^2$. 
In Eq.~(\ref{Xsec}) the kinematical factors are given as usual by
\begin{equation}
  v_L=\left(\frac{\tau}{\kappa^2}\right)^2 
  \quad\quad\quad\quad\quad \mbox{and}
  \quad\quad\quad\quad\quad 
  v_T=\frac{\tau}{2\kappa^2}+\tan^2\frac{\theta_e}{2},
\end{equation}
$\theta_e$ being the electron scattering angle, while $R_L$ ($R_T$) is the
longitudinal (transverse) response function of a nucleus with $Z$ protons
and $N$ neutrons.

The scaling function $F(\kappa,\psi)$ is then defined according to
\begin{equation}
  F(\kappa,\psi) = 
  \frac{\dd^2\sigma / \dd \Omega_e \dd\omega}
     {\sigma_M \left[ v_L G_L(\kappa,\lambda) +
                      v_T G_T(\kappa,\lambda) \right]},
\end{equation}
where the dividing factor, namely the single-nucleon electromagnetic 
cross section, is given in terms of
\begin{subequations}
\label{eq:GLT}
\begin{equation}
  G_L(\kappa,\lambda) =
  \frac{(\kappa^2/\tau)^2
      \left[ {\tilde{G}_E}^2 + \tilde{W}_2 \Delta\right]}
     {2\kappa\left[1+\xi_F(1+\psi^2)/2\right]}
\end{equation}
and
\begin{equation}
  G_T(\kappa,\lambda) =
  \frac{2\tau {\tilde{G}_M}^2 +\tilde{W}_2 \Delta}
      {2\kappa\left[1+\xi_F(1+\psi^2)/2\right]},
\end{equation}
\end{subequations}
with
\begin{subequations}
\begin{eqnarray}
  {\tilde{G}_E}^2 & \equiv & Z {G_{E_p}}^2+N{G_{E_n}}^2 \\
  {\tilde{G}_M}^2 & \equiv & Z {G_{M_p}}^2+N{G_{M_n}}^2 
\end{eqnarray}
\end{subequations}
and
\begin{equation}
  \tilde{W}_2 \equiv \frac{1}{1+\tau} \left({\tilde{G}_E}^2 +
  \tau {\tilde{G}_M}^2 \right).
\end{equation}
In the above the quantity $\psi$, the scaling variable, naturally 
emerges from the RFG studies \cite{Alb88}:
\begin{equation}
  \psi \equiv \frac{1}{\sqrt{\xi_F}} 
     \frac{\lambda-\tau}{\sqrt{(1+\lambda)\tau +
                           \kappa\sqrt{\tau(\tau+1)}}},
\label{eq:psi}
\end{equation}
where $\xi_F\equiv \sqrt{1+\eta_F^2}-1$ is the Fermi kinetic energy in
dimensionless units $(\eta_F\equiv k_F/m_N)$. Any independent pair of
quantities may be used as kinematics variables, $(q,\omega)$, 
$(Q^2,\omega)$, $(\kappa,\lambda)$, $(q,\psi)$, etc., as these are all
functionally related. In studies of QE scaling the last choice proves
useful and in this work we continue to employ this pair. 
Moreover, one finds that a small phenomenological energy shift 
improves the modeling and accordingly we shift 
$\omega\to \omega - E_{\text{shift}}\equiv \omega'$ and correspondingly 
$\lambda\to\lambda'$ and, by substitution of $\lambda'$ for 
$\lambda$ in Eq.~(\ref{eq:psi}), likewise a shifted scaling 
variable $\psi'$:
\begin{equation}
  \psi' \equiv \frac{1}{\sqrt{\xi_F}} 
     \frac{\lambda'-\tau'}{\sqrt{(1+\lambda')\tau' +
                           \kappa\sqrt{\tau'(\tau'+1)}}},
\label{eq:psip}
\end{equation}
where $\tau'=\kappa^2-\lambda'^2$.
Also in the same model a small correction ($\propto \eta_F^2$),
\begin{equation}
  \Delta = \xi_F (1-\psi^2)
  \left[ \frac{\sqrt{\tau(\tau+1)}}{\kappa} +
  \frac{1}{3}\xi_F(1-\psi^2)\frac{\tau}{\kappa^2} \right],
\end{equation}
expresses (see Eqs.~(\ref{eq:GLT})) the impact of the medium on the
single-nucleon electromagnetic cross section~\cite{Amo97}. As far as the 
electric and magnetic form factors of the proton and the 
neutron $G_{E_p}$,$G_{M_p}$, $G_{E_n}$ and $G_{M_n}$ are 
concerned, we adopt the H\"ohler parametrization of Ref.~\cite{Hoe76}. 

The scaling variables $y$, $\psi$ and $\psi^\prime$ have been discussed in
detail in previous work. In particular, the familiar $y$ variable has been
motivated using the impulsive knockout of nucleons from the nucleus as a
starting point (see, for example, Ref.~\cite{Day90}). 
One identifies the allowed region of missing energy and missing momentum for
given $q$ and $\omega$, finding that in the so-called scaling region the lowest
possible value of the missing momentum, where the knockout cross section is
expected to peak, occurs at $-y$. One then argues that at large momentum
transfers the integral of the nuclear spectral function over this kinematic
region ({\it i.e.,} the inclusive cross section insofar as it is dominated by
nucleon knockout) asymptotes to a function that depends only on $y$, but not on
$q$, and hence contains scaling of the first kind. Alternatively, the RFG model
naturally yields a scaling variable, as discussed above. In fact, in this
latter case it not only naturally leads to scaling of the first kind, but, as
stated above, also produces a result that is independent of nuclear species,
namely, scaling of the second kind. This was first suggested in
Ref.~\cite{Alb88} where the concept of superscaling was introduced.
In fact, for typical circumstances involving all but the lightest nuclei, the
various scaling variables are closely related; the inter-relationships have
been explored in depth in Ref.~\cite{Don99-2} and the reader is directed to
that paper for more detail. 

To summarize, scaling of the first kind means that the function
$F(\kappa,\psi)$ loses its dependence upon $\kappa$ for large enough values of
$\kappa$ and naturally occurs in several models.
To get rid as well of the dependence upon the momentum scale set by the 
nuclear structure itself, namely the Fermi momentum $k_F$, it is 
convenient to do as discussed in the Introduction and introduce a 
new dimensionless scaling function $f(\kappa,\psi)$ according to
\begin{equation}
\label{eq:f}
  f(\kappa,\psi) = k_F \, F(\kappa,\psi).
\end{equation}
For the RFG one gets 
\begin{equation}
  f^{\text{RFG}}(\psi) = \frac{3}{4}(1-\psi^2)\theta(1-\psi^2),
\end{equation}
namely, perfect superscaling. More generally, as in previous discussions of
scaling of both the first and second kind, Eq.~(\ref{eq:f}) is the function
upon which we shall focus to ascertain the superscaling properties of the QE
response. 

We shall carry out this task by comparing our predictions with 
the experimental $f(\kappa,\psi)$ extracted from the data of 
SLAC~\cite{Day93} and Jefferson Lab~\cite{Arr99} (see 
Ref.~\cite{Don99-2} for a complete list of high-quality data in 
the revelant kinematic regions); that is, we compare 
the pure RFG response and the 2p-2h excitations induced by the 
MEC currents carried by the pion and the $\Delta$ with the data 
expressed in scaling form. As mentioned earlier, our computation 
will be confined to the 2p-2h transverse response function which 
for the RFG reads \cite{DeP03}:
\begin{eqnarray}
  R_T(\bm{q},\omega) &=& \frac{1}{2} \frac{V}{(2\pi)^9}
    \int \frac{\dd\bm{p}_1\dd\bm{p}_2\dd\bm{p}_1'}
    {16 E_{\bm{p}_1'}E_{\bm{p}_2'}E_{\bm{p}_1}E_{\bm{p}_2}}
    \theta(|\bm{p}_1'|-k_F) \theta(|\bm{p}_2'|-k_F) 
    \theta(k_f-|\bm{p}_1|) 
    \theta(k_F-|\bm{p}_2|) \nonumber \\
  && \times
    \delta[\omega-(E_{\bm{p}_1'}+E_{\bm{p}_2'}-E_{\bm{p}_1}-E_{\bm{p}_2})]
     \sum_{\sigma \tau} \sum_{i,j=1}^3 \, V^4
    \left(\delta_{ij}-\frac{q_i q_j}{\bm{q}^2}\right) \nonumber \\
  && \times 
    \left[{J}_i^\dagger(\bm{p}_1',\bm{p}_1,\bm{p}_2',\bm{p}_2) 
    {J}_j(\bm{p}_1',\bm{p}_1,\bm{p}_2',\bm{p}_2) \right. \nonumber \\
&& \hspace*{4cm} \left.
   - \, {J}_i^\dagger(\bm{p}_1',\bm{p}_1,\bm{p}_2',\bm{p}_2) 
    {J}_j(\bm{p}_1',\bm{p}_2,\bm{p}_2',\bm{p}_1)\right], 
\label{eq:R_T}
\end{eqnarray}
where the two terms in the last factor correspond to direct and 
exchange contributions. In Eq.~(\ref{eq:R_T}) $\bm{p}_1'$ and  
$\bm{p}_2'$ ($\bm{p}_1$ and  $\bm{p}_2$) are the three-momenta 
of the on-shell particles (holes) taking part in the process and 
$E_{\bm{p}}=\sqrt{\bm{p}^2+m_N^2}$.

In Eq.~(\ref{eq:R_T}) one needs the space components of the two-body MEC.
As is well-known there are three of them, namely the pion-in-flight
\begin{subequations}
\label{eq:JMEC}
\begin{equation}
  \label{eq:Jmupi}
  \bm{J}^\mu_f(p_1',p_1,p_2',p_2) = -i \frac{1}{V^2}
    \frac{f_{\pi NN}^2 f_{\gamma\pi\pi}}{\mu_\pi^2}
    (\bm{\tau}^{(1)}\times\bm{\tau}^{(2)})_3
    \Pi(k_1)_{(1)} \Pi(k_2)_{(2)}
    (k_2-k_1)^\mu ,
\end{equation}
the seagull 
\begin{equation}
  \label{eq:Jmus}
  \bm{J}^\mu_s(p_1',p_1,p_2',p_2) = - i \frac{1}{V^2}
    \frac{f_{\pi NN}f_{\gamma\pi NN}}{\mu_\pi^2}
    (\bm{\tau}^{(1)}\times\bm{\tau}^{(2)})_3
    \left[\Pi(k_2)_{(2)}(\gamma^\mu\gamma^5)_{(1)} -
    \Pi(k_1)_{(1)}(\gamma^\mu\gamma^5)_{(2)}\right]
\end{equation}
and the $\Delta$ current (derived here using the Peccei Lagrangian)
\begin{eqnarray}
  \label{eq:JmuD}
  \bm{J}^\mu_\Delta(p_1',p_1,p_2',p_2) &=& -\frac{1}{V^2}
    \frac{f_{\pi NN}f_{\pi N\Delta}f_{\gamma N\Delta}}{2m_N\mu_\pi^2}
    \left\{\left[\left(\frac{2}{3}\tau_3^{(2)} -
    \frac{i}{3}(\bm{\tau}^{(1)}\times\bm{\tau}^{(2)})_3\right)
    \left(j^\mu_{(a)}(p_a,k_2,q)\gamma_5\right)_{(1)}
    \right.\right.\nonumber\\
  && + \left.\left.\left(\frac{2}{3}\tau_3^{(2)} +
    \frac{i}{3}(\bm{\tau}^{(1)}\times\bm{\tau}^{(2)})_3\right)
    \left(\gamma_5 j^\mu_{(b)}(p_b,k_2,q)\right)_{(1)}\right]
    \Pi(k_2)_{(2)} + (1\leftrightarrow2)\right\}. \nonumber \\
\end{eqnarray}
\end{subequations}
In the above $k_1=p_1'-p_1$ and $k_2=p_2'-p_2$ are the momenta of 
the pions entering into each of the 2p-2h MEC diagrams (the 
four-momentum carried by the virtual photon is then $q=-k_1-k_2$) 
and $\mu_\pi$ is the pion mass. Also one has
\begin{equation}
  \Pi(k)_{(i)} = \frac{\bigl(\rlap/k\gamma^5\bigr)_{(i)}}{k^2-\mu_\pi^2},
\end{equation}
the index $(i)$ distiguishing between the two interacting nucleons,
\begin{subequations}
\label{eq:jcurr}
\begin{equation}
  j_{(a)\mu}(p,k,q) = (4 k_{\beta} - \rlap/k\gamma_\beta)
    S^{\beta\gamma}(p,m_\Delta) \frac{1}{2}
    \left(-\gamma_\mu \rlap/q \gamma_\gamma + q_\mu\gamma_\gamma\right)
\end{equation}
and
\begin{equation}
  j_{(b)\mu} (p,k,q) = \frac{1}{2}\left(-\gamma_\beta \rlap/q \gamma_\mu
    + q_\mu\gamma_\beta\right) S^{\beta\gamma}(p,m_\Delta)
    (4 k_{\gamma} - \gamma_\gamma \rlap/k),
\end{equation}
\end{subequations}
where $p_a\equiv p_1-q$, $p_b\equiv p_1'+q$, $m_\Delta$ is the 
$\Delta$ mass and $S^{\beta\gamma}$ the Rarita-Schwinger propagator. 
Finally, $V$ is the volume enclosing our RFG. 

We take $f_{\gamma\pi\pi}=1$, $f_{\gamma\pi NN}=f_{\pi NN}$ and 
$ f_{\pi NN}^2/4\pi=0.08$ for the coupling constants in the Lagrangian.
Although not explicitly indicated above, the currents are meant to include
form factors, both hadronic and electromagnetic. For the former, in 
the present study we assume the standard monopole form
\begin{subequations}
\begin{eqnarray}
  F_{\pi NN}(k^2) &=& \frac{\Lambda_\pi^2-\mu_\pi^2}{\Lambda_\pi^2-k^2}, 
  \\
  F_{\pi N\Delta}(k^2) &=& \frac{\Lambda_{\pi N\Delta}^2}{\Lambda_{\pi
    N\Delta}^2-k^2},
\label{eq:FFD}
\end{eqnarray}
\end{subequations}
with $\Lambda_\pi=1300$~MeV and $\Lambda_{\pi N\Delta}=1150$~MeV, 
while for the latter we use the H\"ohler parameterization for the 
nucleon \cite{Hoe76} and the expression \cite{Dek94}
\begin{equation}
  F_{\gamma N\Delta}(q^2) = \frac{1}{(1-q^2/\Lambda_D^2)^2}
    \left(1-\frac{q^2}{\Lambda_2^2}\right)^{-\frac{1}{2}}
    \left(1-\frac{q^2}{\Lambda_3^2}\right)^{-\frac{1}{2}}
\end{equation}
for the $N\to\Delta$ transition form factor, with
$\Lambda_D^2=0.71$~(GeV/c)$^2$, $\Lambda_2=M+M_\Delta$ and
$\Lambda_3^2=3.5$~(GeV/c)$^2$.  


\section{The $\psi'<0$ region}
\label{sec:psi<0}

\begin{figure}
\includegraphics[clip,width=\textwidth]{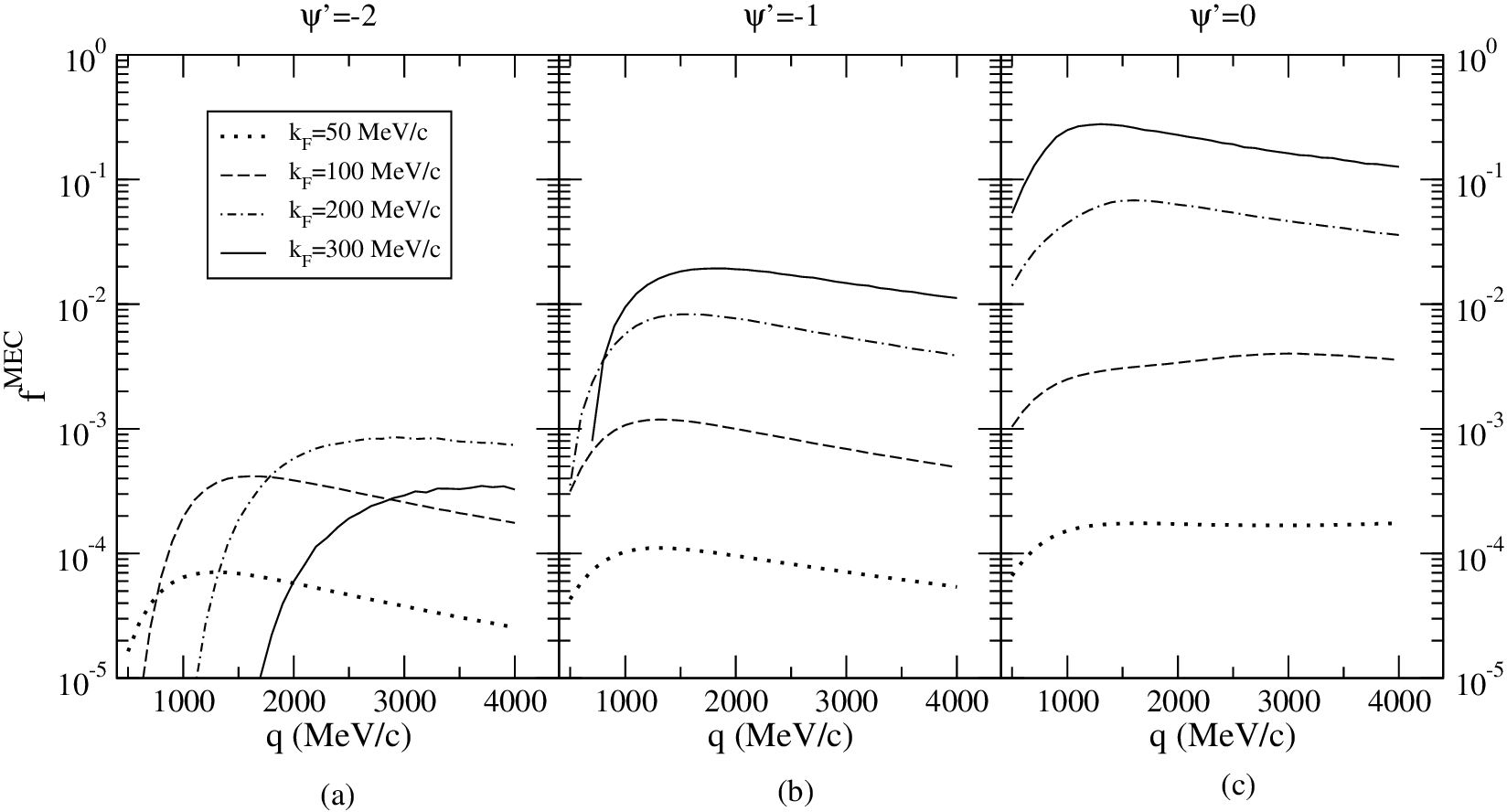}
\caption{\label{fig:psiqle0} MEC scaling function $f^{\text{MEC}}$
versus $q$ for three values of the scaling variable $\psi'$ in the
region below the QEP ($\psi'\leq 0$). The four types of curves
correspond to four choices of $k_F$ ranging from very small (few-body
nuclei) to very large (beyond the heaviest nuclei studied).
}
\end{figure}

Even within the context of the RFG, the 2p-2h excitations induced by 
the MEC currents are nonzero throughout the entire spacelike region 
of the ($\kappa,\lambda$) plane.  In contrast, within the RFG the 
1p-1h contributions are confined to kinematics where 
$-1 \leq \psi \leq +1$, called the 1p-1h response region or Fermi cone. 
In this section we begin by presenting results in the $\psi'<0$ 
region using the RFG for the 2p-2h MEC contributions in the 
scaling function $f$ (referred to as $f^{\text{MEC}}$).
We start by showing in Fig.~\ref{fig:psiqle0} 
the interplay between first- and second-kind scaling by plotting 
$f^{\text{MEC}}$ versus $q$ for a few values of $k_F$, ranging 
from 50 to 300~MeV/c, at $\psi'=-2$, $-1$ and 0. 
While, except for the lightest nuclei, $k_F$ typically lies in 
a much narrower range (200-250 MeV/c), we display results over 
this wider range to illustrate the trend better. In panel a of the 
figure ($\psi'=-2$) one observes that beyond some characteristic 
value of $q$ lying roughly at 2-3 GeV/c first-kind scaling is seen to 
be well fulfilled for typical nuclei.  As nuclei 
within the ``typical nuclei range" become heavier (larger $k_F$) 
this onset is postponed to larger values of $q$. For $\psi'=-1$ 
(panel b), violations of first-kind scaling of $f^{\text{MEC}}$, although 
modest, are clearly apparent and persist for typical nuclei. A 
similar trend is seen for $\psi'=0$ (panel c). Note, also, that very 
light nuclei (for example, deuterium has $k_F\approx 55$ MeV/c) 
would appear to have very little scaling violation in the QEP 
region, since the overall size of $f^{\text{MEC}}$ is 
very minor compared with the usual one-body response for such 
small values of $k_F$.

\begin{figure}
\includegraphics[clip,width=\textwidth]{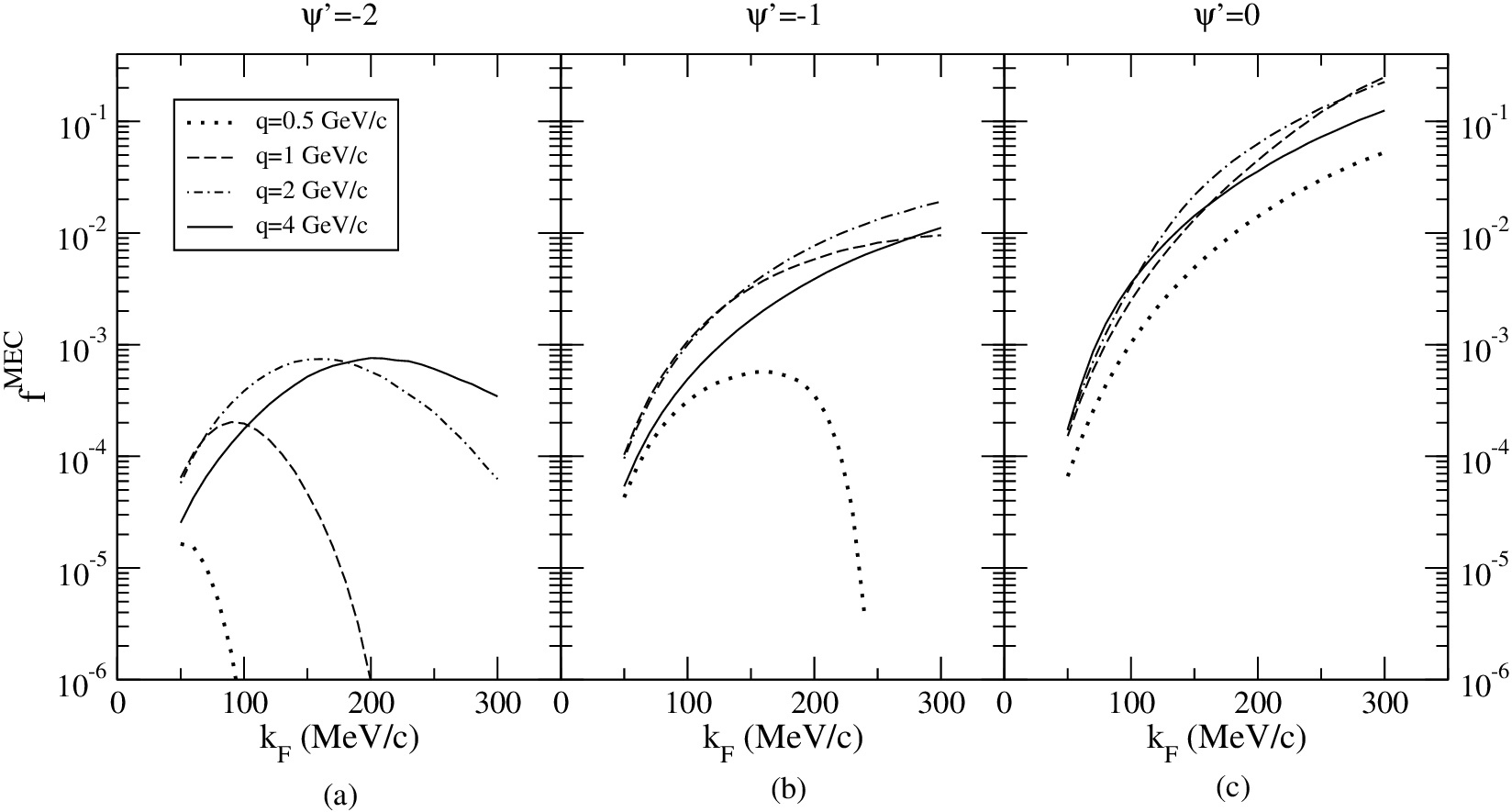}
\caption{\label{fig:psikFle0} As for Fig.~\ref{fig:psiqle0}, 
but now plotting the results as functions of $k_F$ and with
the curves shown corresponding to four choices for $q$.
}
\end{figure}

In context we note that the additional contributions arising from
1p-1h MEC and correlation effects in the region of the QEP also
need to be taken into account --- in the RFG modeling this means
in the region $-1< \psi' < +1$. In Ref.~\cite{Ama03} the net effect
on the scaling behavior of these contributions was explored (see
especially Fig.~7 of that reference) and, as here, scaling violations
of both first and second kinds were predicted. The interference between
the one-body RFG contributions and these additional 1p-1h contributions
reduces the cross section, typically by 10--15\% in the QEP region
for momentum transfers in the few GeV/c range, although there is a 
tendency towards restoration of scaling as $q$ exceeds values 
around 2 GeV/c or so. At the values of $q$ of interest in the present
work scaling of the second-kind is broken roughly proportionally
to $k_F^3$ and hence is relatively unimportant for very light nuclei,
if not for medium-weight ones. For example, in going from carbon to
gold (see below) at $q\approx 1$ GeV/c one finds a second-kind scale
breaking of about 8\% (see~\cite{Ama03}).

In Fig.~\ref{fig:psikFle0} we provide an alternative representation of the
results displayed in Fig.~\ref{fig:psiqle0}, in order to address
directly how the 2p-2h MEC contributions affect the issue of 
second-kind scaling. Here for $\psi'\le0$ we show
$f^{\text{MEC}}$ versus $k_F$ for a few momentum transfers. 
Again, the figure clearly conveys the information that $f^{\text{MEC}}$ 
generally has significant dependence on $k_F$, i.e., does not 
show scaling behavior of the second kind. For typical nuclei 
a range of kinematics can be found where the curves are 
relatively flat; however, for most choices of kinematics 
a strong $k_F$-dependence is observed. For example, in 
the region of the QEP at high $q$ the curves shown in the 
figure go roughly as $k_F^3$ except where $k_F$ is very small.

Note that the behavior found in Fig.~\ref{fig:psikFle0} 
at negative values of $\psi'$ --- namely that $f^{\text{MEC}}$ reaches a
maximum and then decreases with a pronounced $q$-dependence --- is 
just a consequence of the particular kinematical conditions found
in this domain. In fact, at fixed $q$ in the range relevant in 
the present work and negative $\psi'$, when $k_F$ grows one is
actually probing the nuclear response function at lower and 
lower transferred energy $\lambda$ (or $\omega$). This can be 
easily checked by rendering explicit the dependence of $\lambda$ 
on $\psi'$ via Eq.~(\ref{eq:psip}) and by imposing the physical 
requirement $\lambda>0$: one then finds an upper bound on $k_F$, 
specifically, $k_F\alt-q/2\psi'$ (or, in dimensionless units, 
$\eta_F\alt-\kappa/\psi'$).

\begin{figure}
\includegraphics[clip,width=\textwidth]{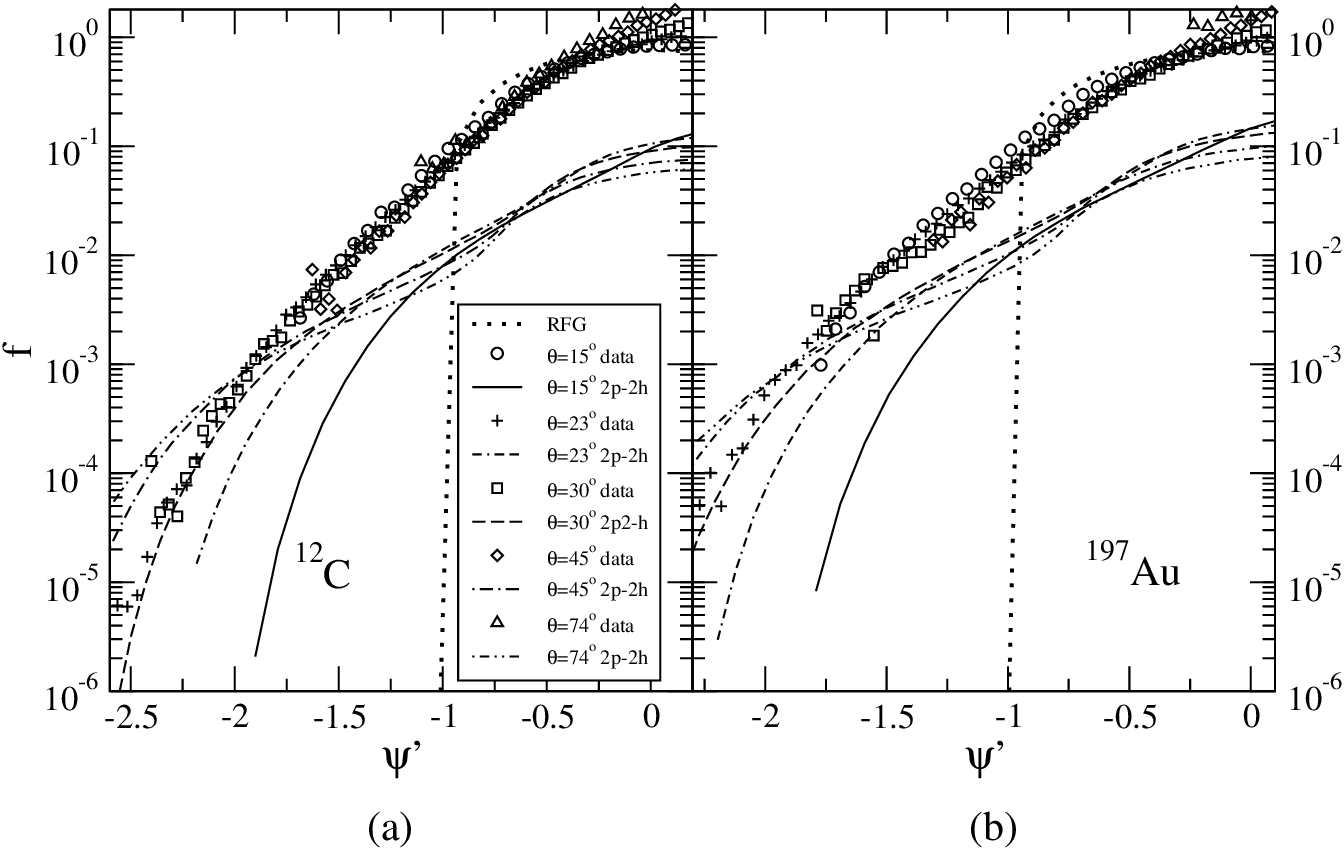}
\caption{\label{fig:CAuJlable0} The dimensionless scaling 
function $f$ plotted versus the scaling variable $\psi'$ 
for JLab data~\protect\cite{Arr99} using carbon (panel a) 
and gold (panel b) targets with electrons of 4045 MeV and 
the scattering angles given in the legend. The legend also labels 
the curves showing $f^{\text{MEC}}$ computed as discussed in the text. 
}
\end{figure}
Next we turn to comparisons of our results with existing high-$q$ data. 
We start by displaying and comparing our results with the data 
obtained at JLab~\cite{Arr99} for $^{12}$C in Fig.~\ref{fig:CAuJlable0}a 
and $^{197}$Au in Fig.~\ref{fig:CAuJlable0}b. The data shown in the 
figure were taken with an electron beam energy of 4045~MeV, scattered 
at angles of $\theta=15^\circ$, $23^\circ$, $30^\circ$, $45^\circ$ and 
$74^\circ$. Note that the momentum transfer $q$ depends not only on 
the beam energy and scattering angle, but also on the energy being 
transferred and therefore on $\psi'$: for the energy transfers 
spanned by the experiment, these values of the scattering angle roughly
correspond to three-momentum transfers in the ranges 1--1.4, 1.6--1.8, 
2--2.4, 2.9--3.1 and 3.9 GeV/c, respectively. Since in this figure we
restrict our attention to the left of the QEP, the actual variation in
momentum transfer is small and, for the sake of the discussion, one 
can just consider the lower limits in the above ranges. As far as
the modeling is concerned, from previous studies~\cite{Mai02} we know to
ascribe a $k_F$ of 228 (245)~MeV/c and an energy shift of 20 (25)~MeV 
to carbon (gold). 

\begin{figure}
\includegraphics[clip,width=\textwidth]{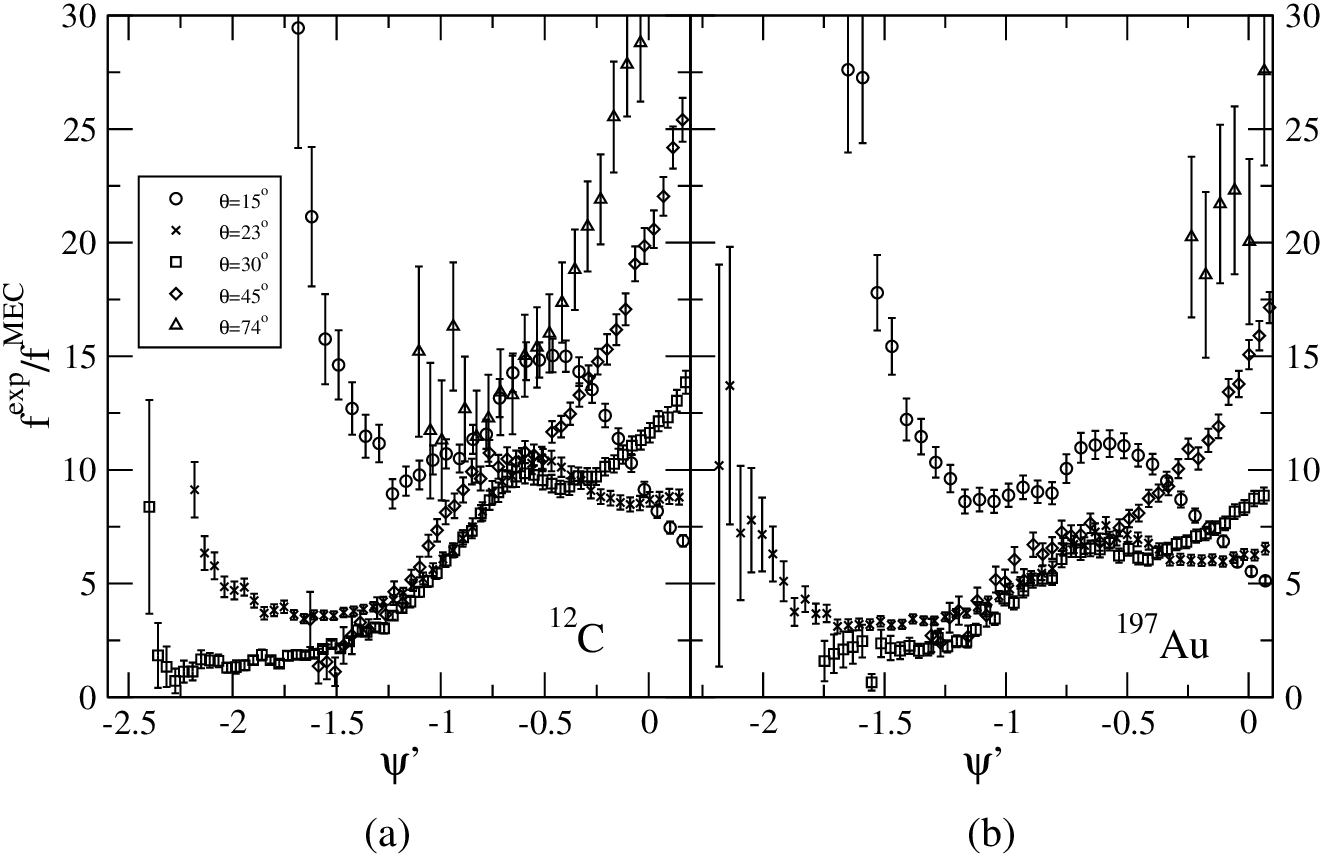}
\caption{\label{fig:CAuJlable0ratio} Ratio of the JLab data 
$f^{\text{exp}}$ shown in Fig.~\ref{fig:CAuJlable0} to the 
calculated $f^{\text{MEC}}$ both for carbon (panel a) and gold (panel b).
}
\end{figure}

Comparisons with the data must be made carefully: each data set 
with a given scattering angle (and therefore roughly a given 
value of $q$, as noted above) extends down only to a specific 
value of $\psi'$. In particular, for $\theta=15^\circ$, $23^\circ$, 
$30^\circ$, $45^\circ$ and $74^\circ$, for carbon (panel a in the 
figure) these values are $\psi'_{min}\approx -1.7$, $-2.6$, $-2.4$, 
$-1.6$ and $-1.1$, respectively. For gold (panel b) the corresponding 
numbers are $\psi'_{min}\approx -1.8$, $-2.3$, $-1.7$, $-1.3$ and $-1.3$, 
respectively. 
This is made clearer in Fig.~\ref{fig:CAuJlable0ratio} where the
ratio of the JLab data $f^{\text{exp}}$ given in 
Fig.~\ref{fig:CAuJlable0} to the calculated $f^{\text{MEC}}$ 
is shown both for carbon (panel a) and gold (panel b). Clearly
one sees from this representation that typically the 
size of the MEC contributions is 10\% or less (i.e., the ratio in
Fig.~\ref{fig:CAuJlable0ratio} is 10 or larger), except at large
negative values of $\psi'$. Such conditions can be met by these data sets
only for $\theta=23^\circ$ and $30^\circ$, and in the latter case
really only for carbon, as the $30^\circ$ gold data do not extend
below about $-1.7$. In the discussions to follow, this will limit the
most stringent test of second-kind scaling to the $23^\circ$ case.

Let us begin the comparisons of $f^{\text{MEC}}$ with the data in
the negative $\psi'$ region by addressing the issue of 
{\em scaling of the 
first kind}, namely the behavior of $f$ versus $\psi'$ as 
the momentum transfer $q$ is varied. From Fig.~\ref{fig:CAuJlable0} 
one sees that outside of the Fermi cone the 2p-2h scaling 
function is small at low momentum transfers
(forward angles) and then grows with $q$. For a specific choice
of $\psi'$ there appears to be a characteristic value of $q$ at
which the first-kind scaling behavior sets in (see also 
Fig.~\ref{fig:psiqle0} above). For example, if one fixes one's 
attention to $\psi'=-2$, this stabilization occurs for $q$ a 
little larger than 2 GeV/c. Ideally, comparisons would be made
with the data for a wide range of momentum transfers at large
negative $\psi'$ (say, below $-2$) where the 2p-2h MEC effects are
predicted to be relatively large and thereby would see the approach
to first-kind scaling. However, as is made clear by
Fig.~\ref{fig:CAuJlable0ratio}, for very large $q$ the range 
in $\psi'$ is relatively
narrow and only the $23^\circ$ and $30^\circ$ data extend below
$\psi'=-2$, and there only for carbon. In this region of kinematics
to test high-$q$, low-$\psi'$ first-kind scaling behavior
one therefore has only the carbon data for $\theta=23^\circ$ where,
as noted above, $q\approx 1.6$ GeV/c and for $\theta=30^\circ$ where
$q\approx 2$ GeV/c. From the figures one sees that the 
amount of scaling violation
provided by the 2p-2h MEC contributions is clearly compatible with
the scaling violations seen in the data. At less negative values
of $\psi'$ the ratios presented in Fig.~\ref{fig:CAuJlable0ratio} 
show that the relative size of $f^{\text{MEC}}$ is typically only
about 10\% or less of the full $f$ and that the first-kind scaling
violations, even at the highest values of $q$, are again compatible
with the amount seen in the data.

\begin{figure}
\includegraphics[clip,width=0.9\textwidth]{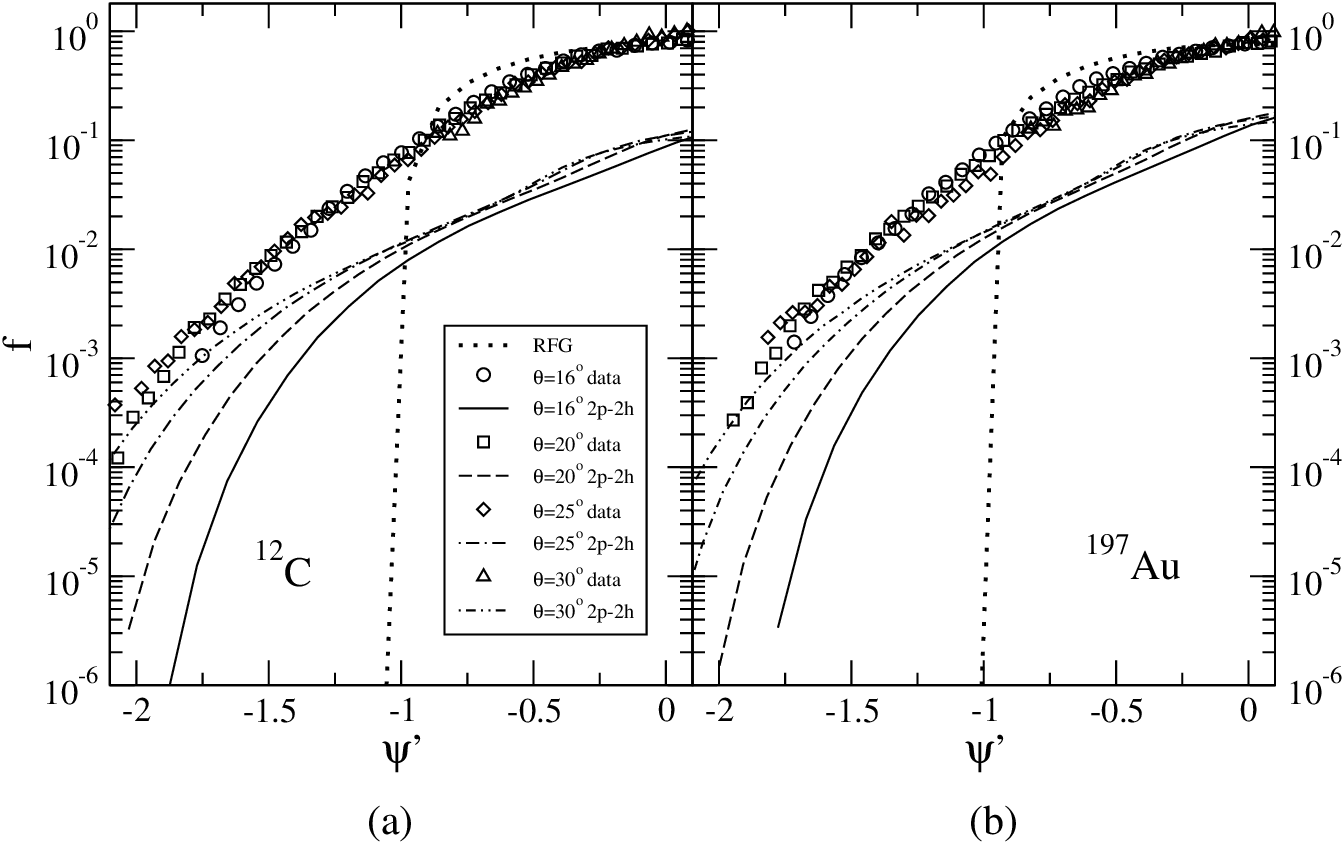}
\caption{\label{fig:CAuSlacle0} As for Fig.~\ref{fig:CAuJlable0}, but
now for SLAC data~\protect\cite{Day93} taken at 3595 MeV for the 
angles listed in the legend. The curves are as in 
Fig.~\ref{fig:CAuJlable0}.
}
\end{figure}
\begin{figure}
\includegraphics[clip,width=0.9\textwidth]{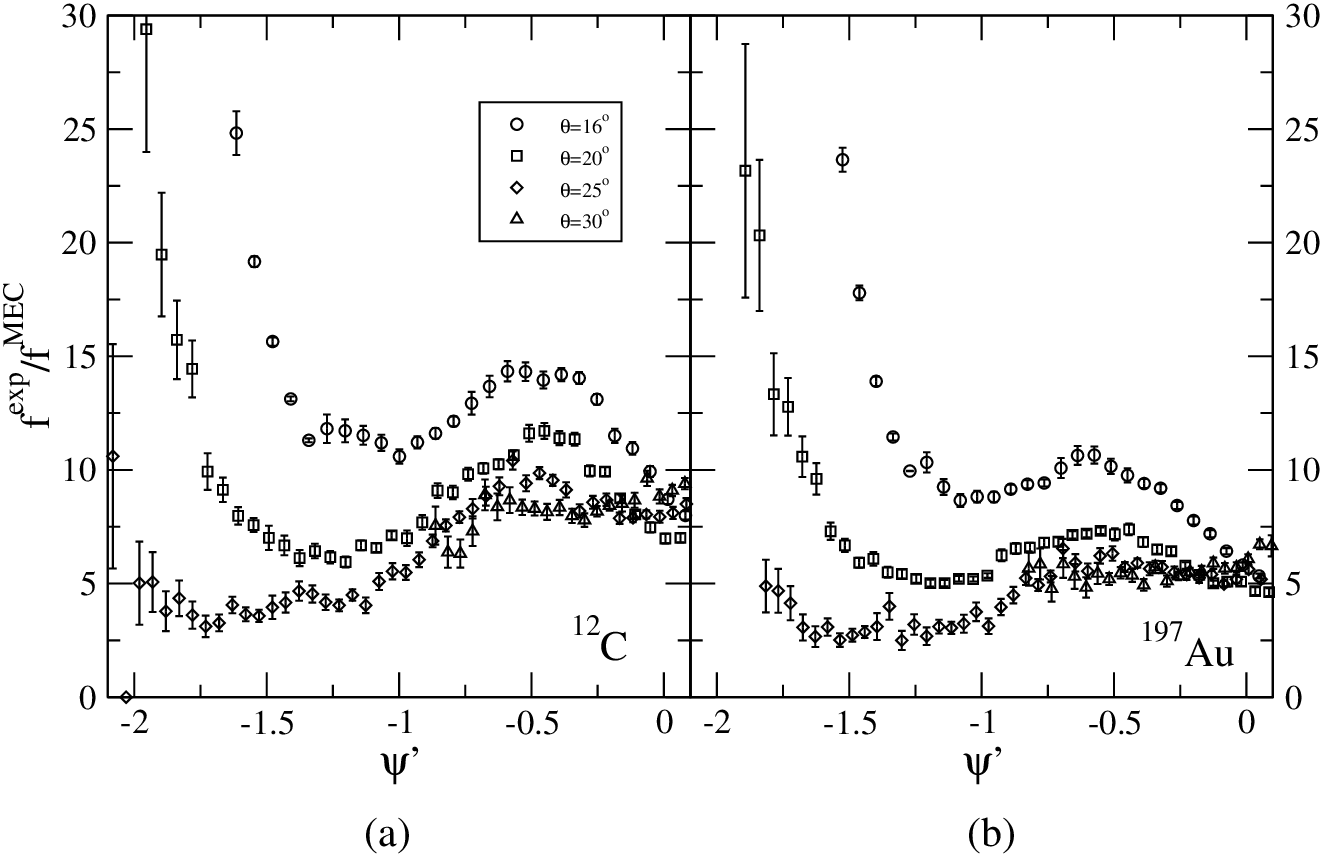}
\caption{\label{fig:CAuSlacle0ratio} Ratio of the SLAC data 
$f^{\text{exp}}$ shown in Fig.~\ref{fig:CAuSlacle0} to the 
calculated $f^{\text{MEC}}$ both for carbon (panel a) and gold (panel b).
}
\end{figure}
As discussed above, note that inside the Fermi cone, where the 
2p-2h MEC provide roughly a 10\% contribution, other competing effects 
must also be taken into account. In particular, it is known that 
the 1p-1h MEC and correlation contributions must be considered.
These interfere with the one-body contributions --- the basic
RFG contributions --- and so decrease 
the scaling function from its RFG value in this domain 
by roughly the same amount as the 2p-2h MEC provide as additions 
\cite{Ama02a,Ama02b,Ama03}. Since the 1p-1h and 
2p-2h response functions add incoherently, one expects them
to cancel each other to a large extent in the QEP region. 

For the sake of completeness, in Figs.~\ref{fig:CAuSlacle0} and 
\ref{fig:CAuSlacle0ratio} we display the scaling function 
$f^{\text{MEC}}$ for $^{12}$C and $^{197}$Au as in 
Figs.~\ref{fig:CAuJlable0} and \ref{fig:CAuJlable0ratio}, but now here 
for data from 
SLAC~\cite{Day93} taken with a beam of energy 3595~MeV and scattering 
angles $\theta=16^\circ$, $20^\circ$, $25^\circ$ and $30^\circ$. 
Accordingly, the associated momentum transfers probed in these 
experiments now span the ranges 1--1.7, 1.2--1.4, 1.5--1.7 and 
1.9--2.4~GeV/c, respectively. Again, the actual momentum range
probed in the negative-$\psi'$ region is smaller, and the lower 
limits may be taken as typical values. The minimum values attained 
for the scaling variable are, for carbon $\psi'_{min}\approx -1.8$, 
$-2.1$, $-2.1$ and $-0.9$, and for gold $\psi'_{min}\approx -1.7$, 
$-2.0$, $-1.8$ and $-0.8$, for the above set of angles, respectively. 
The results observed are rather similar to those presented in 
Figs.~\ref{fig:CAuJlable0} and \ref{fig:CAuJlable0ratio}.

\begin{figure}
\includegraphics[clip,width=\textwidth]{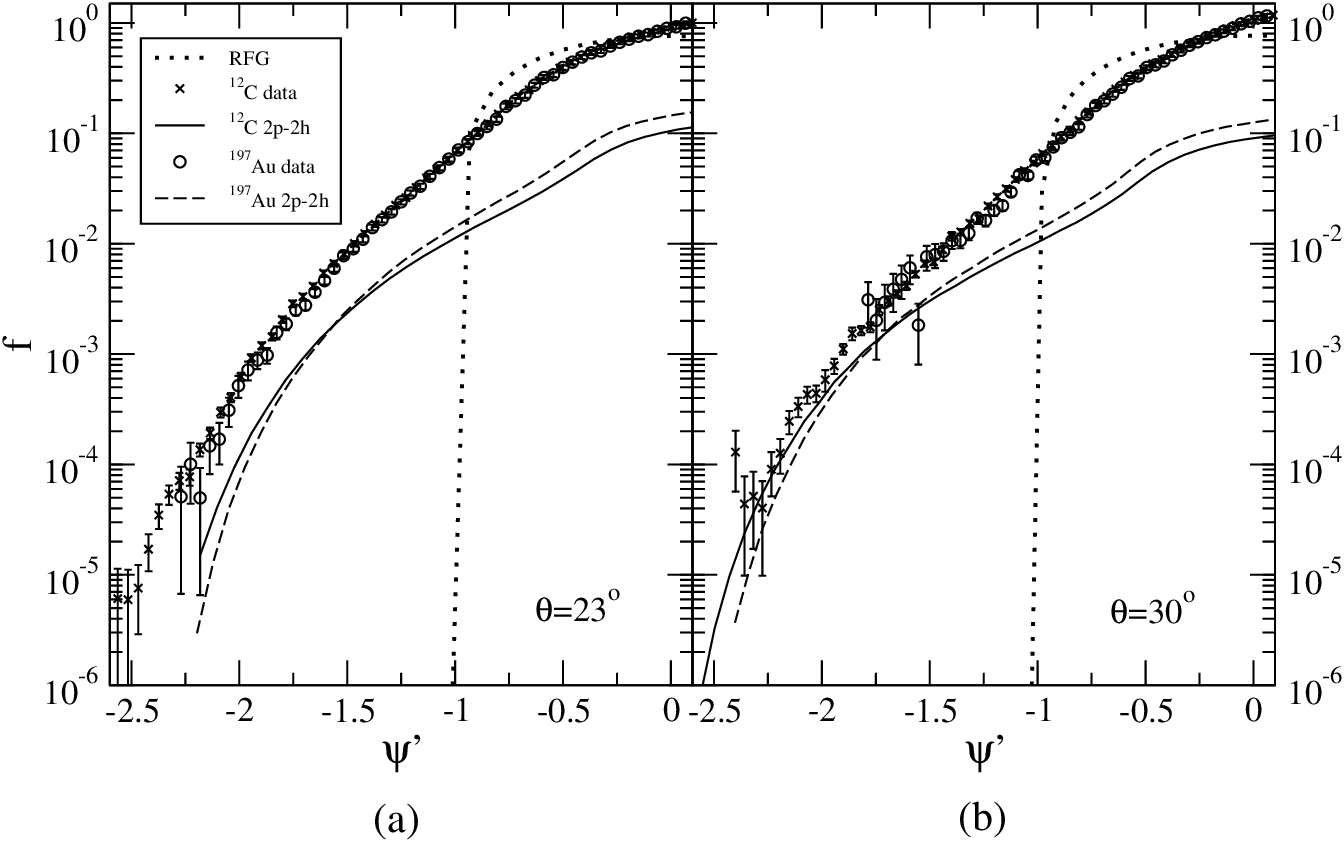}
\caption{\label{fig:CAuJlab2330le0} The dimensionless scaling 
function $f$ plotted versus the scaling variable $\psi'$ for JLab 
data~\protect\cite{Arr99} at $\theta=23^\circ$ (panel a) and 
$30^\circ$ (panel b), 
as well as the results obtained for $f^{\text{MEC}}$ for carbon 
(crosses and solid curve) and gold (circles and dashed curve).
}
\end{figure}
Let us next discuss the {\em second-kind scaling} behavior seen at
large negative $\psi'$.  In fact, as the above figures show,
the only cases where both carbon and gold data extend down to
this region are those for $\theta=23^\circ$ and $30^\circ$.
Accordingly, in Fig.~\ref{fig:CAuJlab2330le0} the JLab data for 
$\theta=23^\circ$ (panel a) and $30^\circ$ (panel b) are separated
out from Fig.~\ref{fig:CAuJlable0} and again compared with 
$f^{\text{MEC}}$, but now simultaneously for carbon 
(crosses and solid curve) and gold (circles and dashed curve).
Starting with the $23^\circ$ case shown in 
Fig.~\ref{fig:CAuJlab2330le0}a, one observes the 
following: at the QEP $(\psi'=0)$ one finds that $f^{\text{MEC}}$ 
is about 13\% of the experimental $f$ and that the second-kind scaling 
violation implied by the difference between the curves for carbon and gold
amounts to about 4\%. As already noted above, the results shown here do 
not take into account the 1p-1h MEC + correlation contributions which
decrease the overall non-impulse-approximation effect, making these
numbers over-estimates. Both classes of contributions go roughly as
$k_F^3$ at high $q$ and so the partial cancellation is relatively
independent of the actual species of nuclei being examined.

In contrast, at large negative $\psi'$ 
the only scale-breaking contributions in the present model are
those provided by the 2p-2h MEC terms and, in fact, on the one
hand we see that they can have relatively large effects when
compared with the full experimental scaling function --- for instance, 
at $\psi'=-2$ one finds that $f^{\text{MEC}}$ is about
16\% of the experimental $f$. On the other hand, the second-kind 
scaling violation implied by the difference between the curves 
for carbon and gold amounts to about 7\% for these kinematics, 
interestingly with the gold curve lying below that for carbon, 
in contrast to the behavior above $\psi'\approx -1.6$ where the 
reverse is true. Thus, for the high-$q$, large-negative-$\psi'$ 
region covered by the $23^\circ$ data for carbon and gold one 
sees some violation of scaling of the second kind, although such MEC 
effects alone appear not to destroy the excellent scaling 
seen in the existing data. Indeed, slight adjustments in the 
choices of $k_F$ and $E_{\text{shift}}$ can be made (these are, 
after all, empirically determined~\cite{Mai02}) and thereby can 
also provide changes of this size. However, as shown in
Fig.~\ref{fig:CAuJlab2330le0}b where results at $30^\circ$ and hence
large $q$ are shown, it does not take too much for the MEC effects,
including the amount of second-kind scale breaking they provide, to
become significant. Unfortunately, as noted above, the gold data
do not extend down to very low values of $\psi'$ for these conditions
and so it is not presently possible to see clear evidence for such
contributions. Finally, we note that for intermediate negative 
values of $\psi'$ the net degree 
of second-kind scale-breaking is expected to be only a few percent 
and clearly compatible with what is seen in the data. 


\section{The $\psi'>0$ region}
\label{sec:psi>0}

In the region $\psi'>0$ at sufficiently high $q$ baryon 
resonances, meson production and eventually deep inelastic scattering (DIS)
make significant, even dominant, contributions to the total 
response. This regime has recently been explored in 
Ref.~\cite{Bar04} in which an {\em inelastic} version of 
the basic RFG approach, together with an extension called 
the ERFG, were studied (the reader is directed to the cited 
reference for details). Additionally, here we also have 
contributions from $f^{\text{MEC}}$.

\begin{figure}
\includegraphics[clip,width=\textwidth]{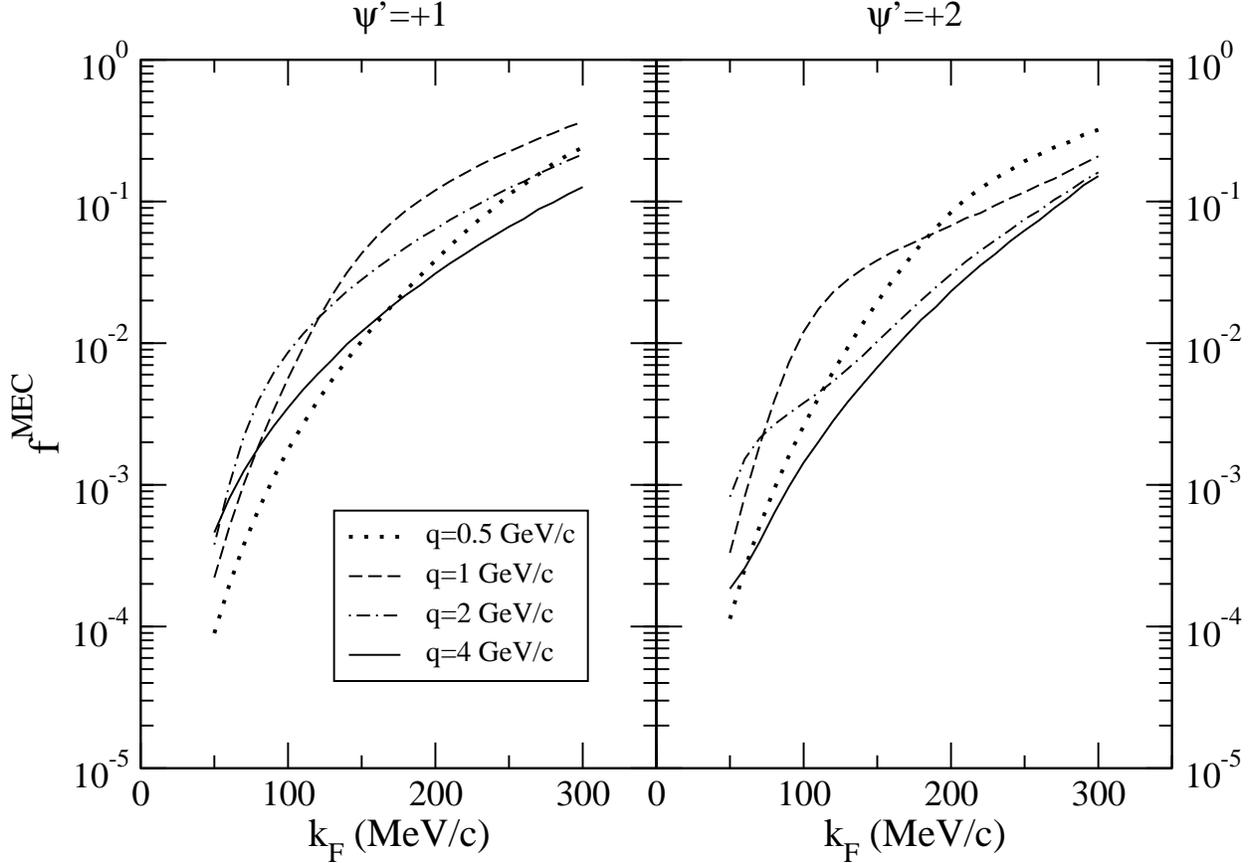}
\caption{\label{fig:psikFgt0} As for Fig.~\ref{fig:psikFle0}, but now for
the region above the QEP.
}
\end{figure}
To explore how $f^{\text{MEC}}$ may or may not violate second-kind 
scaling in the domain $\psi'>0$, as in the previous section we begin 
by showing results versus $k_F$ in Fig.~\ref{fig:psikFgt0} 
at $\psi'=1$ and 2. Here it is quite apparent that the 
2p-2h MEC excitations alone substantially break the second-kind 
scaling. For example, at $\psi'=1$ in going from $^{12}$C to $^{197}$Au
there is roughly a 20\% increase in $f^{\text{MEC}}$ at $q=1$~GeV/c and a
35\% increase at $q=4$~GeV/c.

\begin{figure}
\includegraphics[clip,width=0.9\textwidth]{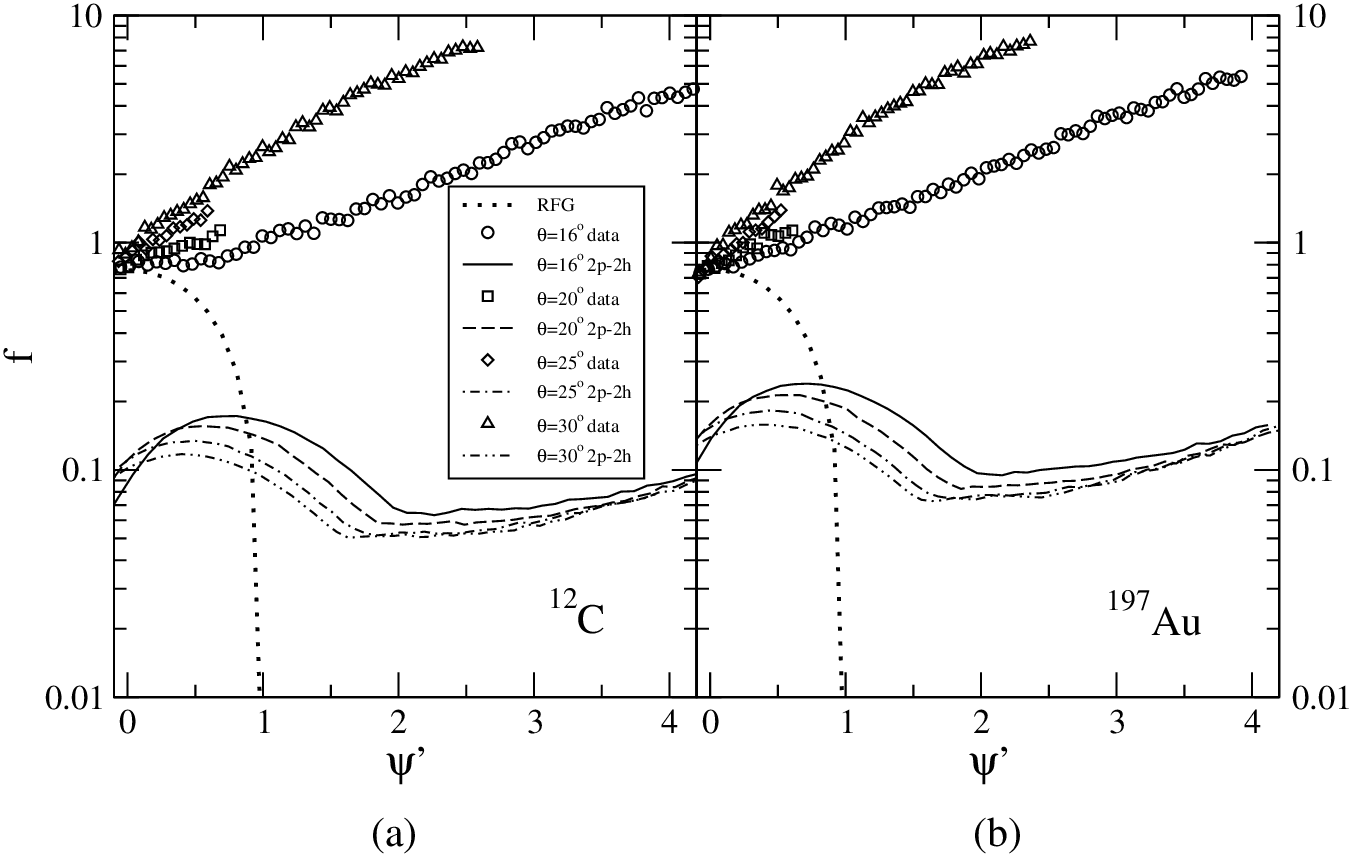}
\caption{\label{fig:CAuSlacgt0} As for Fig.~\ref{fig:CAuSlacle0}, 
showing data from SLAC~\protect\cite{Day93} and results for 
$f^{\text{MEC}}$, but now for the region above the QEP ($\psi'\geq 0$). 
}
\end{figure}
\begin{figure}
\includegraphics[clip,width=0.9\textwidth]{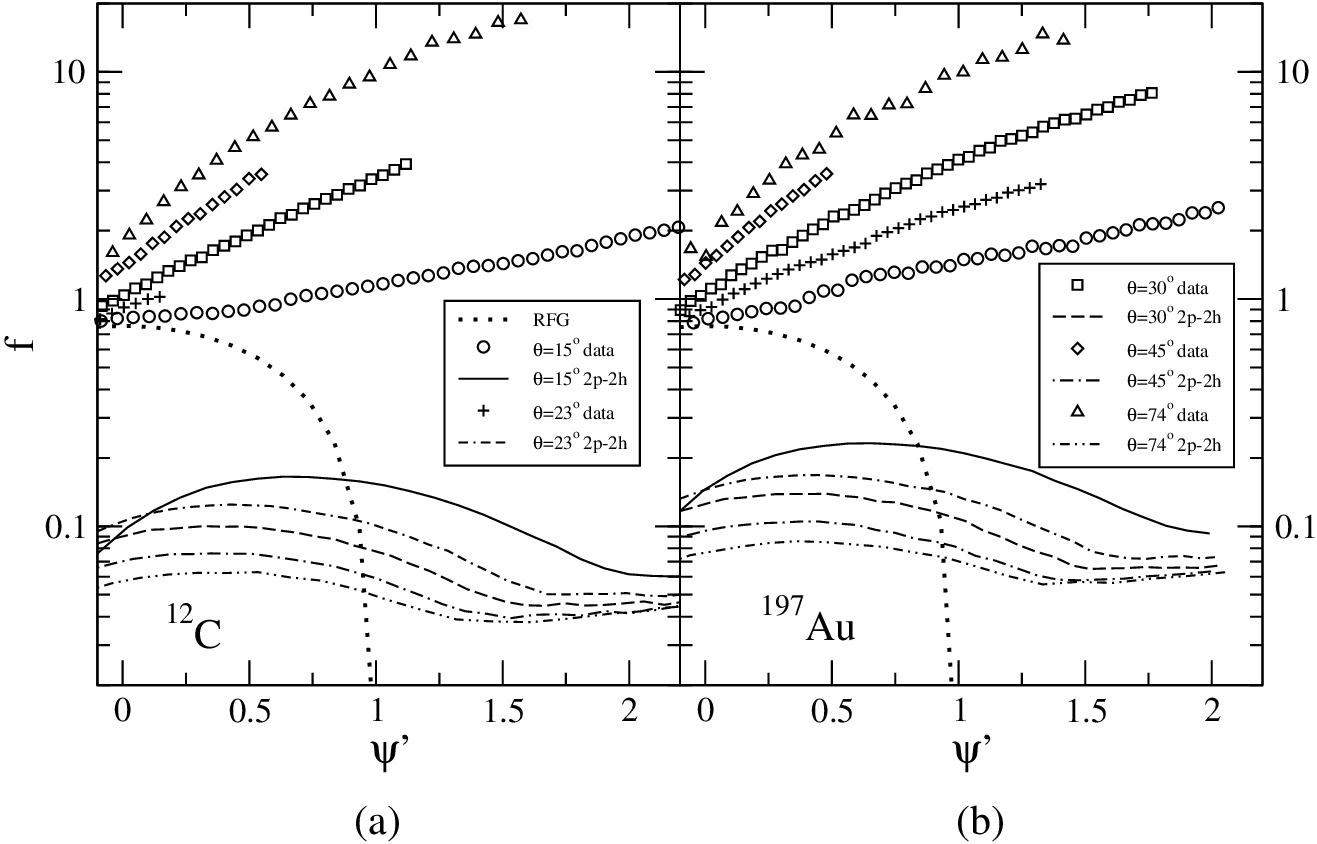}
\caption{\label{fig:CAuJlabgt0} As for Fig.~\ref{fig:CAuSlacgt0}, 
but now showing comparisons with data from JLab~\protect\cite{Arr99}.
}
\end{figure}

Furthermore, the 2p-2h MEC contributions 
that arise from the modeling discussed in the present work can provide
significant contributions to the total scaling function, especially at
lower values of $q$. To get a feeling for
the size of these when $\psi'$ is positive, we display 
$f^{\text{MEC}}$ together with $f^{\text{RFG}}$ in 
Fig.~\ref{fig:CAuSlacgt0} (where the SLAC data are also shown) 
and in Fig.~\ref{fig:CAuJlabgt0} (where the JLab data are 
displayed), again for $^{12}$C and $^{197}$Au. 
In the figures it is seen that $f^{\text{MEC}}$ is significant around
$\psi'=1$, accounting for roughly 10--20\% of the data for $q$ below
2~GeV/c, down to a few percent at the highest momenta. Note that
the trend is in the right direction: the heavier nucleus 
(larger $k_F$) has a larger 2p-2h MEC contribution than does the
lighter one, in concert with what is observed in the data. 

Putting these two pieces of information together we see that the
degree of second-kind scaling violation that arises from the 2p-2h
MEC contributions obtained in the present work is a few percent at
modest $q$, dropping to the sub-one-percent level at higher $q$.
This is smaller than the amount found in Ref.~\cite{Mai02}, namely
that in the domain $\psi'>0$ second-kind scaling in the total 
scaling function $f$ is violated at about the 20\% level. There it was 
observed that the inelasticity mentioned above itself violates
scaling of the second kind and clearly dominates the total inclusive
cross section. The follow-up study in Ref.~\cite{Bar04} makes this
point even clearer (see, for example, Fig.~12 in that reference where
the $\theta=16^\circ$ SLAC data also shown in Fig.~\ref{fig:CAuSlacgt0} 
are compared with the inelastic RFG and ERFG model results).
Thus, at least at the point reached in addressing
the nature of the inclusive cross section in the region above the QEP,
we find that, while the 2p-2h MEC effects can be important at modest
values of $q$, they do not apparently break the second-kind scaling
behavior by more than a few percent and that the actual scale-breaking
observed comes from other effects.

A final thing to note is that 
the 2p-2h MEC contributions violate first-kind scaling in the opposite 
direction to the trend displayed by the data, since one observes 
a decreasing of $f^{\text{MEC}}$ with increasing $q$. This is 
also opposite to the behavior seen in the scale-breaking seen 
in the inelastic modeling reported in Ref.~\cite{Bar04}, which 
behaves more the way the data do. At higher values of $\psi'$, 
$f^{\text{MEC}}$ stays relatively constant, whereas the data 
definitely grow, the more so the higher the value of $q$.
 

\section{Role of the $\Delta$}
\label{sec:Delta}

\begin{figure}
\includegraphics[clip,width=\textwidth]{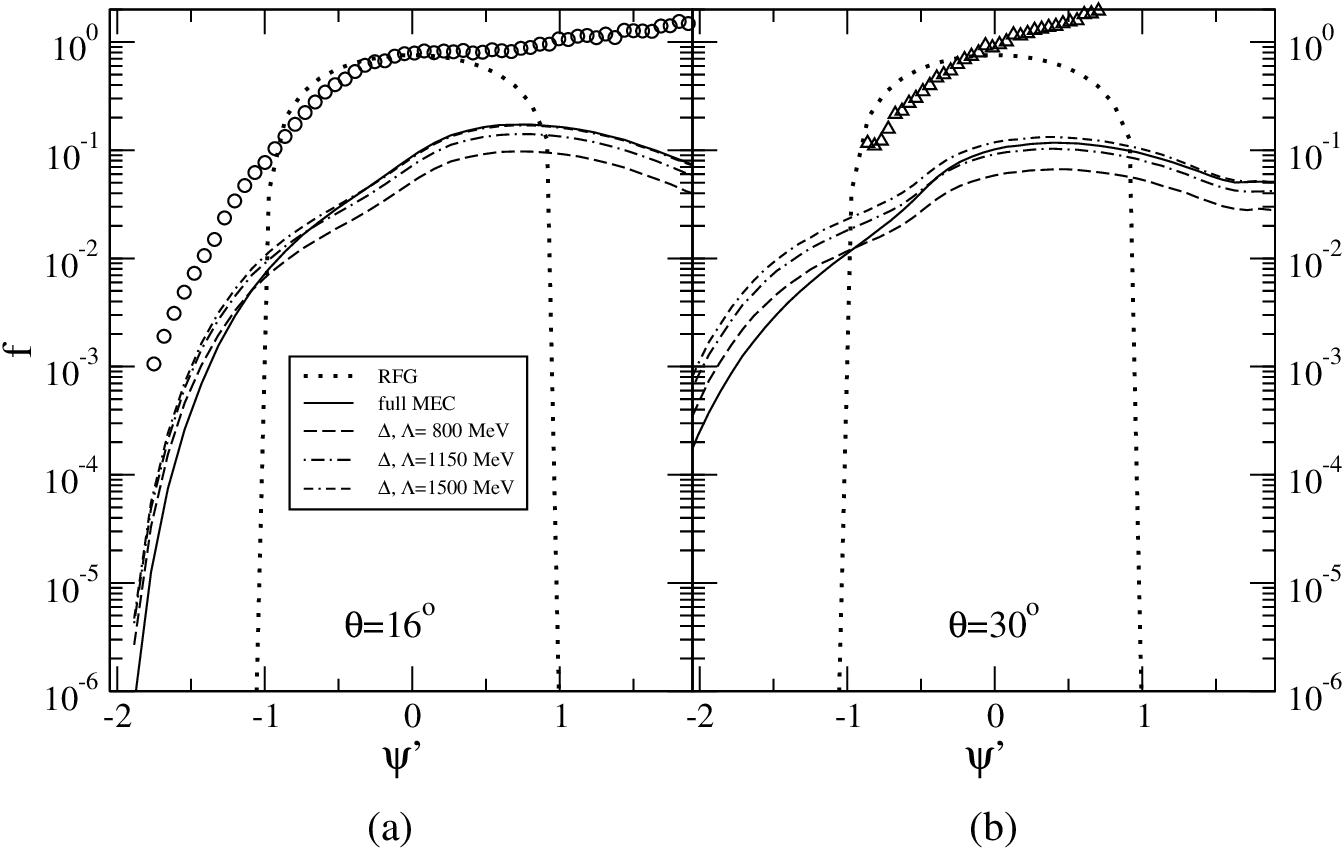}
\caption{\label{fig:CSlacD} The scaling function for $^{12}$C at the kinematics
  of the experiments of Ref.~\protect\cite{Day93} for $\theta=16^\circ$ (a) and
  $30^\circ$ (b). The dotted line represents $f^{\text{RFG}}$ and the solid
  line $f^{\text{MEC}}$ in the full model, whereas all the other curves
  represent $f^{\text{MEC}}$ including only the $\Delta$-current for different
  choices of the strong $\pi N\Delta$ cutoff,
$\Lambda_{\pi N\Delta}$ (abbreviated to $\Lambda$ here).
Results are shown for $\Lambda =800$~MeV (dash),
  1150~MeV (dot-dash) and 1500~MeV (dot-dash-dash). As usual, for
carbon one has $k_F=228$~MeV/c
  and $E_{\text{shift}}=20$~MeV.
}
\end{figure}
\begin{figure}
\includegraphics[clip,width=\textwidth]{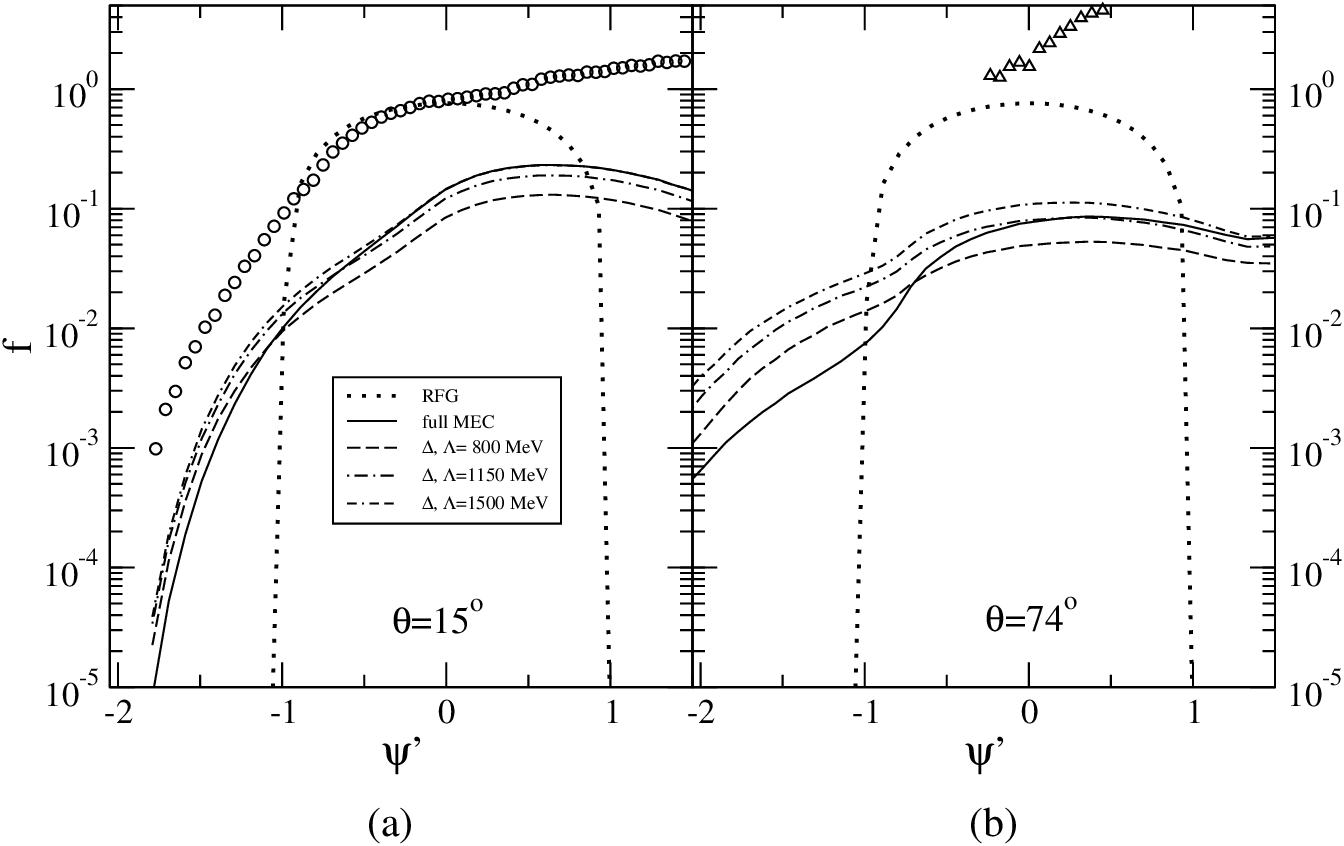}
\caption{\label{fig:AuJlabD} As in Fig.~\protect\ref{fig:CSlacD}, but for
  $^{197}$Au at the kinematics of the experiments of
  Ref.~\protect\cite{Arr99} for $\theta=15^\circ$ (a) and $74^\circ$ (b);
  $k_F=245$~MeV/c and $E_{\text{shift}}=25$~MeV.
}
\end{figure}
$\Delta$-dominance in both the 1p-1h and 2p-2h sectors of the 
RFG Hilbert space has been plainly demonstrated in Refs.~\cite{Ama02a} 
and \cite{DeP03}, respectively. As we shall see, this comes about 
through a rather delicate cancellation, and accordingly we shall
explore in a bit more depth two aspects of these issues.
On the one hand, we wish to investigate the sensitivity of the 
contributions made by the $\Delta$-current to the 2p-2h excitations 
with respect to variations of the poorest-known parameter entering 
into the definition of the $\Delta$-current itself, namely, the
cutoff in the hadronic form factor in Eq.~(\ref{eq:FFD}). On the 
other hand, we wish to gauge the relative role played by the 
$\Delta$ via comparisons of the 2p-2h MEC contribution computed 
using the whole set of MEC in Eqs.~(\ref{eq:JMEC})  with those arising
from the $\Delta$ alone.

In Fig.~\ref{fig:CSlacD} the basic RFG scaling function and 
$f^{\text{MEC}}$ are displayed, together with the SLAC data 
taken at $\theta=16^\circ$ (panel a) and $\theta=30^\circ$ 
(panel b), versus $\psi'$ for $^{12}$C. The 2p-2h MEC scaling 
function is computed in the full model and with the
$\Delta$ alone,  in the latter case in three different versions, 
corresponding to three values of the cutoff in Eq.~(\ref{eq:FFD}), 
namely $\Lambda_{\pi N\Delta}=800$, $1150$ and $1500$~MeV.
The same analysis is carried out for $^{197}$Au and presented in 
Fig.~\ref{fig:AuJlabD}, now with the JLab data at $\theta=15^\circ$ 
(panel a) and $\theta=74^\circ$ (panel b). Interesting features 
emerge from the two figures.

First, concerning $^{12}$C, we recall the discussions of kinematics
made earlier: in panel (a) one has $1\alt q\alt1.7$~GeV/c, whereas 
in panel (b) one has $1.9\alt q\alt2.4$~GeV/c. For these ranges of 
three-momentum transfer it is observed that within the Fermi cone 
the non-$\Delta$ terms (those arising from diagrams having only
nucleons and pions) play quite a minor role, almost insignificant 
at small $q$, and that the harder the $\pi N\Delta$ 
vertex, the larger the $\Delta$ contribution. In contrast, in the
scaling region ($\psi'\alt-1$),  it turns out that the $\Delta$ 
by itself yields a contribution that is substantially larger than the
one found with the full MEC and that this becomes more pronounced 
as $q$ grows.

This behavior, in fact, stems from the interferences between the
$\Delta$ and non-$\Delta$ contributions which are of opposite 
sign~\cite{DeP03}. That is, we see that the non-$\Delta$ currents, 
while giving rise to small contributions in themselves, actually 
open the door to significant interference effects. This is 
emphasized in Fig.~\ref{fig:AuJlabD}, where the
MEC are probed at substantially larger momenta, namely 
$q\approx3.9$~GeV/c in panel (b). 


\section{Discussion and Conclusions}
\label{sec:concl}

The concept of scaling is based on having some basic ``elementary''
cross section for scattering from the constituents in a system
which can reasonably be factored out, leaving a reduced cross section
whose nature is determined essentially by the distribution of the
constituents. Given that a decomposition of this sort is indeed 
reasonable, that is, that a factorization of this type occurs at
least for the dominant contributions in the cross section (coherent
scatterings, for example, would not tend to have such a property),
it is often the case that the reduced cross section is in fact a
relatively universal function of one or more scaling variables,
namely, that the results scale and one has a very compact representation 
of the cross section \cite{Don99-1}. 
This is certainly true when the constituents of the system 
are ``simple'' or behave as such, as happens for the partons 
in the nucleon (asymptotic freedom) or for the electrons in 
an atom at appropriate energies. However, it can remain true 
even when the constituents are complex, as for atoms in 
solids and, as the studies discussed in the present work indicate, 
for nuclei at high $q$ and excitations
spanning a wide range from below the quasielastic peak through
the resonance region and into the regime where DIS is beginning
to become applicable. On the other hand, it is certainly not a 
valid way to proceed in all circumstances; for instance, very near
threshold in electroexcitation one has collectivity and the
factorization is clearly broken.

Even in the kinematical region where the dominant contribution appears
to scale, one knows that other processes can also play some role and 
therefore that several questions arise. Do such additional contributions 
show either first- or second-kind scaling behaviors or not? If not, are
they sizable contributions relative to the dominant (scaling)
processes, and therefore, if they do not scale, is the modeling in 
conflict with observation or not? Or, said another way, how much 
scaling violation is to be expected from these additional contributions?

In the present context, where we are focusing on the scaling, QEP and 
resonance regions for inclusive electron scattering at high $q$, one can 
think of several physical processes of this type that might provide 
some degree of scale-breaking beyond the basic impulsive process (the 
RFG model in the present work) which does have the scaling properties. In 
particular, we know from work already done~\cite{Ama02a,Ama02b,Ama03} 
that MEC effects and correlations in the 1p-1h sector
do not in general have scaling of either the first or second kind,
although the degree of scale-breaking is relatively small under
most circumstances. Moreover, in the resonance region and beyond
we know that inelastic, but impulsive scattering from nucleons
in the nucleus also breaks both kinds of scaling~\cite{Mai02,Bar04}, 
and in a way that tends to improve the agreement with the data (which 
also do not scale in that region).

Still other effects remain to be explored. In particular, it would
be highly desirable to have relativistic treatments of correlation
effects not just for the 1p-1h sector alluded to above, but also 
in the 2p-2h sector, since these are known to be sizable at least 
in non-relativistic calculations 
at low momenta~\cite{Alb84}; one would like relativistic modeling
of short-range NN correlations as well, as they are expected
to contribute especially in the large-negative-$\psi'$  
region~\cite{Ben94}; furthermore, additional many-body interaction 
effects (such as via RPA correlations) may play a role; and nuclear 
binding effects arising, for example, in a relativized shell model 
approach should also be explored. Most of these are 
presently being revisited in a relativistic context in other on-going 
work.

In the present work we have considered one other specific process
which naturally does not scale, namely, the effects arising from 2p-2h MEC
contributions. 
We employ currents arising both from diagrams containing the $\Delta$
and those built only from nucleons and pions (non-$\Delta$ terms), and
in a brief section present results to show the delicate interplay
between these two classes of diagrams and to indicate the 
cutoff-dependence in the former. Our present studies are based 
on recent work~\cite{DeP03} where these pieces of the cross section 
were derived and comparisons were made with other existing studies.

Our motivation for the present work is a specific one: here we have 
focused on the scaling 
properties of the 2p-2h MEC contributions. Results are presented for
kinematics ranging from far below the QEP (at large negative
$\psi'$), in the region of the QEP (near $\psi'=0$) and beyond
(at positive $\psi'$) where resonances and DIS can both play significant
roles. We have found that scaling of both the first and second kinds
is generally broken by these contributions. 

At negative $\psi'$ we find potentially large contributions
from 2p-2h MEC effects --- these may be comparable to the total
response at very large negative $\psi'$ and amount typically to
about 10\% at $\psi'=-1$. When $\psi'$ is very negative
we have seen a stabilization with increasing $q$ at fixed $\psi'$ set
in, so that at very high values of $q$ the first-kind scaling 
behavior appears to be recovered. On the other hand, also at large 
negative $\psi'$, we see violations of second-kind scaling that are
compatible with the existing data in this region; however, the
results obtained here suggest that, were data on several nuclei to
become available at still higher values of $q$, one might begin to
see significant second-kind scaling violations. 

In the vicinity of the QEP ($-1<\psi'<+1$) we have continued to see
a delicate interplay between the MEC+correlation effects in the
1p-1h sector and the present 2p-2h MEC effects --- the former
interfere with the impulsive RFG response and lead to a decrease
in the net 1p-1h result, typically by about 10\%, 
whereas the 2p-2h MEC effects go in the
opposite direction by roughly the same amount. The net result,
at least when 1-body, 1p-1h (MEC+correlation) and 2p-2h (MEC only)
contributions are taken into account, is a tendency 
to restore the final total answer back to where one began. 
One should be cautious, however, in drawing conclusions that
are too strong at this point, since relativistic 2p-2h correlation 
contributions are as yet not available.

In the large-positive-$\psi'$ region we find that the second-kind
scaling is again violated, but not by very much compared with
the data. Indeed, it is clear from previous work that the main
scaling violations in this regime stem from the inelastic, but 
impulsive excitations. This is not to say that the 2p-2h MEC
contributions are insignificant --- they account for perhaps 10--20\%
of the total strength in the resonance region --- just that they
appear to provide only few percent violations of second-kind scaling. 
Interestingly, in this region the first-kind scaling is also
violated by the 2p-2h MEC contributions, but with a trend that is
opposite to what is observed in the data and also in the inelastic
modeling. In summary, in the resonance region we find that the net result
is typically about 80--90\% from inelastic 1-body contributions
({\it i.e.,} in the extended RFG model of \cite{Bar04}), to which must be
added the roughly 10--20\% from 2p-2h MEC effects. 


\section{Acknowledgments}
This work has been supported in part by the INFN-MIT Bruno Rossi Exchange
Program and by MURST (contract No.~2001024324\_007) and in part (TWD) by funds
provided by the U.S. Department of Energy under cooperative research agreement 
No.~DE-FC02-94ER40818.

\end{document}